\documentclass[twocolumn,prc,aps,floatfix]{revtex4}
\usepackage[dvips]{graphicx}
\begin{document}
\title{Enhancement of variation of
fundamental constants in ultracold atom and molecule systems near
Feshbach resonances}
\author{Cheng Chin}
\affiliation{James Franck Institute and Physics Department, The
University of Chicago, IL 60637}
\author{V.V. Flambaum}
\affiliation{School of Physics, The University of New South Wales,
Sydney NSW 2052, Australia}
\date{\today}

\begin{abstract}
Scattering length, which can be measured in Bose-Einstein condensate
and Feshbach molecule experiments, is extremely sensitive to the
variation of fundamental constants, in particular, the
electron-to-proton mass ratio ($m_e/m_p$ or $m_e/\Lambda_{QCD}$,
where $\Lambda_{QCD}$ is the QCD scale). Based on single- and
two-channel scattering model, we show how the variation of the mass
ratio propagates to the scattering length. Our results suggest that
variation of $m_e/m_p$ on the level of $10^{-11}\sim 10^{-14}$ can
be detected near a narrow magnetic or an optical Feshbach resonance
by monitoring the scattering length on the $1\%$ level. Derived
formulae may also be used to estimate the isotopic shift of the
scattering length.
\end{abstract}

\maketitle


  Theories unifying gravity with other interactions suggest a possibility
of temporal and spatial variation of the fundamental constants of nature
(see e.g. review \cite{Uzan} where the theoretical models and
results of measurements are presented).
There are  hints for the variation of the
fine structure constant, $\alpha=e^2/\hbar c$,
strength constant of the strong interaction and fundamental masses
in Big Bang nucleosynthesis, quasar absorption spectra and Oklo
natural nuclear reactor data
(see e.g.\cite{var,DFW,Lam,Ivanchik}) .
However, a majority
of publications report only limits on  possible variations
 (see e.g. recent review \cite{karshenboim}).

The hypothetical unification of all interactions implies that variation
of the electromagnetic interaction constant $\alpha$ should be accompanied
by the variation of masses and the strong interaction constant.
 Specific predictions need a model. For example, the grand unification
model  discussed in \cite{Langacker:2001td} predicts that
 the
quantum chromodynamic (QCD) scale  $\Lambda_{QCD}$
 (defined as the position of the Landau pole in the logarithm for the
running strong coupling constant)
is modified as follows: $\delta \Lambda_{QCD} / \Lambda_{QCD}
\approx 34 \delta \alpha / \alpha$.
The variation of quark and electron masses in this model is  given by
$\delta m / m \sim 70 \delta \alpha / \alpha $.
This gives an estimate for the variation of the dimensionless ratio
\begin{equation} \label{mQCD}
{\delta(m/ \Lambda_{QCD}) \over(m/\Lambda_{QCD})}\sim 35 {\delta \alpha
\over \alpha}
\end{equation}
The proton mass is proportional to $\Lambda_{QCD}$
 ($m_p \approx 3\Lambda_{QCD}$). Therefore, we have
\begin{equation} \label{memp}
{\delta(m_e/m_p) \over(m_e/m_p)}\sim 35 {\delta \alpha
\over \alpha}
\end{equation}
This result is strongly model-dependent.
However, the large coefficients in these expressions are generic for
 grand unification models, in which modifications come from high energy scales:
they appear because the running strong coupling constant and
 Higgs constants (related to mass) run faster than $\alpha$.
 This means that if these models
are correct the variation
of masses and strong interaction may be easier to detect than the variation
of $\alpha$.

One can measure only variation of the
dimensionless quantities, therefore we want to extract from the measurements
 variation
of the dimensionless ratio $m_e/m_p$.
A number of limits on variation of $m_e/m_p$ and  $m_q/\Lambda_{QCD}$
( where $m_q$ is the light current quark
mass with the dependence on the normalization point removed)
have been obtained recently from consideration
of Big Bang Nucleosynthesis, quasar absorption spectra
and Oklo natural nuclear reactor which was active about
1.8 billion years ago \cite{FS,oliv,dmitriev,FS1,DFW,Pana,Lam,Ivanchik}.
The present day limits have been obtained from the atomic clocks
experiments. The  most recent results have been presented in \cite{tedesco},
see also original measurements in \cite{atomic}.

In this letter, we propose a new approach to measure the variation
of electron-to-proton mass ratio $\beta=m_e/m_p$ with extremely high
sensitivity. We show that observables in atomic Bose-Einstein
condensates (BECs) and Feshbach molecules experiments can be
extremely sensitive to $\beta$. This dependence can be understood as
the molecular potential is essentially electronic, while the atomic
mass is essentially baryonic. The high sensitivity results from the
fact that the scattering length depends critically on the energy of
the last bound state, while the variation of mass-ratio scales the
overall vibrational energy, which lead to an stronger accumulative
effect on the higher-lying bound states.

Recent experiments have well characterized the atomic scattering
length based on the mean-field shifts in BECs, coherent
atom-molecule coupling \cite{rb85precision}, and photo-association
\cite{wynar2000}, Feshbach \cite{chin04}, and radio-frequency
\cite{bartenstein05} spectroscopy. In these works, scattering
lengths are typically determined to $\sim 1\%$ in the BEC-type
experiments and to $10^{-4}$ in the molecular spectroscopy
experiments. In this paper, we theoretically evaluate the
enhancement factor $K$, defined as the ratio of the fractional
change of the scattering length to that of $\beta$, namely, $\delta
a/a=K(\delta\beta/\beta)$. Our results show that huge enhancement
factors of $K=10^9\sim10^{12}$ can be obtained near narrow Feshbach
resonances. This results can provide alternative stringent and
independent tests on the variation of fundamental constants.

To quantify the dependence of scattering length on fundamental
constants, we first consider a single-channel scattering model
described by molecular potential  $V(r)$ which at large distances
approaches the van der Waals interaction $V(r)=-C_6/r^6$. Here $r$
is the atomic separation and $C_6$ is the van der Waals coefficient.
This potential very well characterizes the interaction of ultracold
neutral atoms. Atomic $s$-wave scattering length $a$ in this
potential is given by \cite{flambaum93}

\begin{eqnarray}\label{a}
a=\bar{a}[1-\tan(\phi-\pi/8)],
\end{eqnarray}

\noindent where $\bar{a}=c(2\mu C_6/\hbar^2)^{1/4}$
 is the mean scattering length, $c=2^{-3/2}\Gamma(3/4)/\Gamma(5/4)\approx0.47799$
is a constant, $\mu$ is the reduced mass of the two interacting
atoms and $\phi$ is the scattering phase shift.

For most atoms, the associated molecular potential can support a
large number of vibrational bound states and one can calculate the
scattering phase shift semi-classically as \cite{flambaum93}

\begin{eqnarray}\label{phi}
\phi=\int_{r_i}^\infty \hbar^{-1}\sqrt{-2\mu V(r)} dr,
\end{eqnarray}

\noindent where $r_i$ is the classical inner turning point at zero
energy. The number of vibrational bound state is $N=[\phi/\pi+3/8]$
\cite{flambaum93}, where $[x]$ is the integer part of $x$.

Equation~(\ref{a}) suggests that fundamental constants can influence
the scattering length in two ways, either by varying the mean
scattering length or by varying the scattering phase shift. Varying
Eq.~(\ref{a}), we derive

\begin{eqnarray}\label{d_a}
\frac{\delta
a}{a}=\frac{\delta\bar{a}}{\bar{a}}-\frac{(a-\bar{a})^2+\bar{a}^2}{a\bar{a}}\delta\phi.
\end{eqnarray}

 The scattering length has dimension of length,
 therefore, we have to specify units we will use.
We are only interested in enhanced effects of the variation. In this
case  choice of units is not important since their variation can be
neglected anyway. Note that the mean scattering length in
Eq.~(\ref{a}) can be presented as  $\bar{a}=const.$~$\beta^{-1/4}
a_B$ where
 $a_B$ is the Bohr radius. The phase $\phi$ may be presented as
$\phi=(A_N/\beta)^{1/2}\int \sqrt{-V(r)/E_H} d(r/a_B)$ where
$E_H=\hbar^2/m_e a_B^2$ is the Hartree energy,
$A_N$ is the number of nucleons in the nucleus. Therefore, it is
convenient to use atomic units. We  normalize all lengths to $a_B$
and energies to  $E_H$.
 Performing variation of
 $\bar{a}$ and  $\phi$
 we obtain that in atomic units they depend on variation of
 the electron-to-proton mass ratio
$\beta$ only:

\begin{eqnarray}\label{d_abar}
\frac{\delta\bar{a}}{\bar{a}}=-\frac14\frac{\delta\beta}{\beta},\,\,\,\,\,\frac{\delta\phi}{\phi}=-\frac12\frac{\delta\beta}{\beta}.
\end{eqnarray}

The variation of $r_i$ disappears since the classical momentum
$p=\sqrt{-2\mu V(r)}$ is zero at the turning point. In the
derivation of the above equations, we have used $\delta({V/E_H})=0$.
Indeed, the molecular potential is determined by the Coulomb
interaction between electrons and nuclei. Therefore, in the
non-relativistic limit all parameters of this potential are
proportional to the atomic units, there are no other natural units
in this problem.
 The relativistic corrections
 bring some additional weak dependence
on  $\alpha$ (see e.g.  \cite{alpha}) which may be neglected here.
Indeed, the relativistic corrections are of the order of $Z^2
\alpha^2$ and relatively small.  For Cs $Z^2 \alpha^2=0.16$, for Li
$Z^2 \alpha^2=0.0005$ (here $Z$ is the nuclear charge).

 Combining Eqs.~(\ref{d_a}) and (\ref{d_abar}) and assuming the potential
supports many bound states ($\phi\approx N\pi\gg \pi$), we obtain
\begin{eqnarray}\label{d_a_final}
\frac{\delta a}{a}=\frac{N\pi}2
\frac{(a-\bar{a})^2+\bar{a}^2}{a\bar{a}}\frac{\delta\beta}{\beta}=K\frac{\delta\beta}{\beta},
\end{eqnarray}

\noindent where $K$ is the desired enhancement factor.

 Note that this equation may also be used to estimate the
isotopic shift of the scattering length if it is small
($\delta\beta/\beta=(A_N-A_N')/A_N$).

Two enhancement factors are identified here: one from the number of
bound states $N$ and one from the resonant behavior of the
scattering length $((a-\bar{a})^2+\bar{a}^2)/(a\bar{a})$. Note that
we have large enhancement here in two limiting cases: very large $a$
and very small $a$.

The first enhancement factor is larger for heavier atoms with deeper
molecular potentials as well as higher density of states. For
example, singlet potential of Cs$_2$ supports $N\approx150$
vibrational states, while that of Li$_2$ only has $N=38$ states.

More remarkably, when the scattering length is large compared the
mean scattering length, the second enhancement factor approaches
$a/\bar{a}$, which can be very large near a potential resonance. In
this limit, the binding energy of the last molecular state is simply
given by
 $E_b=\hbar^2/2\mu a^2$ (for $a>0$). We can then write the
variation of the binding energy in the threshold regime as

\begin{equation}\label{d_E_final}
\frac{\delta
E_b}{E_b}=-N\pi\frac{(a-\bar{a})^2}{a\bar{a}}\frac{\delta\beta}{\beta}.
\end{equation}

Atoms at low temperatures constitute an ideal system to measure the
scattering length since the scattering length is the only parameter
which characterizes the interaction properties. In the case of
cesium, the scattering length in the singlet potential has been
determined to be $a=280.37$~$a_B$ with a relative accuracy of $\sim
10^{-4}$ \cite{chin04}. Based on the number of Cs$_2$ vibrational
states $N\approx 150$ and the mean scattering length of
$\bar{a}=95$~$a_B$, Eq.~(\ref{d_a_final}) suggests an overall
enhancement factor of
 $K\sim 400$.

Similar sensitivity of $\delta a/a=2\times 10^{-4}$ on the singlet
molecular potential of lithium-6 atoms is also obtained recently
\cite{bartenstein05}. Due to the relatively fewer molecular number
of $N=38$ and the small singlet scattering length of
$a=$45.17~$a_B$, the enhancement factor is only $K=62$.

For real atoms, the molecular potential is multi-channel in nature.
This opens up new possibilities to modify the scattering lengths and
to obtain higher enhancement factors. In particular, when a bound
state in a closed channel is tuned close to the scattering state in
the open channel, Feshbach coupling between the two channels can
lead to a resonant enhancement of the scattering length according to

\begin{eqnarray}\label{A}
A=a(1+\frac{\Delta E}{E_o-E_m}),
\end{eqnarray}

\noindent where $A$ is the scattering length near the resonance, $a$
is the off-resonant scattering length, $\Delta E$ characterizes the
Feshbach coupling strength, and $E_m$ is the energy
of the bound state, which is supported by the closed channel.
We assume that the long-range asymptotic of the
 the closed channel potential
is $V_c=E_c-C_6/r^6$
 and the asymptotic of  open channel
potential is $V_o=E_o-C_6/r^6$. Note that  the Feshbach resonance
may have a finite decay width $\Gamma_m$, therefore, the resonance
energy
 denominator is better described as $E_o-E_m+i\Gamma_m/2$.
 However, the elastic width vanishes at zero kinetic energy of colliding
particles and inelastic width is small. To simplify our
consideration we will neglect  $\Gamma_m$ and other possible
nonlinear effects in the channel coupling.

To determine the dependence of $A$ on fundamental constants, we
first assume that the molecular potentials of both channels and the
Feshbach coupling are proportional to the atomic unit of energy, namely,

\begin{equation}
\delta \frac{E_c}{E_H}=\delta \frac{E_o}{E_H} = \delta \frac{\Delta
E}{E_H}=0
\end{equation}
This assumption needs corrections if $\Delta E$ contains the
hyperfine interaction component ( see the variation of the hyperfine
interaction in \cite{tedesco}). However, near the resonance the weak
dependence of $\Delta E$ on fundamental constants is relatively
insignificant and may be neglected.

Variation of Eq.~(\ref{A}) then gives

\begin{equation}\label{d_A}
\frac{\delta A}A=\frac{\delta a}a+\frac{A-a}A\frac{\delta
E_m}{E_o-E_m}.
\end{equation}

We note that the energy of the bound state $E_m$ depends on the
vibrational energy and therefore on the mass ratio.
 To estimate this dependence, we assume the molecular potential $V_c(r)$ supports
many bound states, and we may use a WKB approximation to estimate
the molecular energy $E_m$ using equation

\begin{equation}\label{M}
\phi_M=\int_{r_i}^{r_o} \hbar^{-1}\sqrt{2\mu(E_m-V_c(r))} dr=M\pi
\end{equation}

\noindent where the integer $M$ is the vibrational quantum number of
the bound state and $r_i(r_o)$ is the classical inner(outer) turning
point for a total energy of $E_m$.

Equation (\ref{M}) relates $E_m$ to the atomic mass $m$. Varying Eq.
(\ref{M}), we can relate the level shift to the variation of the
mass ratio $\beta$. The result is

\begin{equation}\label{d_M}
\delta E_m=-\frac{\phi_M}{2\pi\rho(E_m)}\frac{\delta\beta}{\beta}
\end{equation}

\noindent where $\rho(E_m)=1/D$ is the density of states at the
energy $E_m$ and can be estimated from the energy splitting $D$
between adjacent vibrational levels.

Combining Eqs.~(\ref{d_A}-\ref{d_M}) and using $\phi_M=M\pi\gg\pi$,
we show that near a Feshbach resonance, the change of scattering
length is dominated by the resonance term and can be written as

\begin{equation}\label{d_A_final1}
\frac{\delta A}{A}=\frac{M}2\frac{(A-a)^2}{Aa}\frac1{\rho(E_m)\Delta
E}\frac{\delta\beta}{\beta}.
\end{equation}

There are three enhancement factors in Eq.~(\ref{d_A_final1}): the
first one is the vibrational quantum number of the closed channel
bound state $M$; the second one depends on the scattering length and
approaches $A/a$ near the resonance. These two factors resemble
those from the single channel calculation, see
Eq.~(\ref{d_a_final}). With the Feshbach resonance, however, one is
given the flexibility to adjust the second enhancement factor. The
third factor of $(\rho(E_m)\Delta E)^{-1}=D/\Delta E$ in Eq.~(14) is
a new effect. It shows that higher energy resolution can be obtained
by using a narrower Feshbach resonance and larger level shifts occur
when the closed channel has a larger vibrational state energy
splitting $D$.






In the following, we take ultracold Li and Cs atoms as two examples
to estimate the dependence of the molecular binding energy or the
scattering length on the mass ratio. For both atomic species,
multiple Feshbach resonances exist and the vibrational energy
splittings in the highest singlet channel near the lowest triplet
scattering threshold are $D=h$~10$\sim$20~GHz \cite{julienne}. We
will adopt $\rho=D^{-1}=(h$~10~GHz$)^{-1}$ in the following
estimation.

For $^6$Li, a broad resonance at 834~G with a width of 300~G
($\Delta E$=h~840~MHz) and a background scattering length of
$a=$-1405~$a_B$ \cite{bartenstein05} can be a candidate to reach
very large scattering lengths. This resonance is induced by a closed
channel bound state with $M=38$ and molecular binding energy has
been measured to $\sim 1\%$ near this resonance
\cite{bartenstein05}. This measurement is performed deep in the
quantum threshold regime with scattering lengths up to
$A\sim1300$~$a_B$. Here, we can rewrite Eq.~(\ref{d_A_final1}) in
terms of the threshold binding energy
 $E_b=\hbar^2/2 \mu A^2$ as

\begin{equation}\label{d_E_final2}
\frac{\delta E_b}{E_b}=-M\frac{(A-a)^2}{Aa}\frac1{\rho(E_m)\Delta
E}\frac{\delta\beta}{\beta}.
\end{equation}

Based on the Li parameters, we calculate the total enhancement to be
$K=-38\times (-4)\times 12=1824$, which is indeed much larger than
the single channel result we obtained. The $1\%$ binding energy
measurement thus translates into a mass ratio measurement with an
accuracy of $5\times 10^{-6}$.

Feshbach molecules of $^6$Li are also observed near a narrow
resonance at 546~G with a width of $0.1$~G ($\Delta E$=h~0.28~MHz)
\cite{huletnarrow}. This resonance is induced by the same
vibrational manifold with $M=38$ and the background scattering
length is $a=62$~$a_B$. A measurement similar to
\cite{bartenstein05} should be possible at $A=620$~$a_B$, where we
expect the enhancement factor to be as large as $K=-38\times 8.1
\times 36,000\approx-10^7$.

An extreme case would be the narrow $g$-wave Feshbach resonances
observed in $^{133}$Cs, which are induced by molecular states with
four units of angular momentum. These resonances have a typical
resonance width of $\sim 5$~mG ($\Delta E$=h~3.5~kHz)
\cite{herbig03} and can be induced by states in the closed singlet
channel with $M\sim 140$. Since the threshold regimes of these
resonances are expected to be too narrow to access experimentally,
we return to Eq.~(\ref{d_A_final1}) and calculate the variation of
the scattering length. Assuming a scattering length measurement is
performed at $A=1000$~$a_B$ near a narrow resonance and the
background scattering length is $160$~$a_B$, we find the overall
enhancement factor is as large as $K=70\times (4.8) \times
(3\times10^6)\approx10^9$.

High resolution measurement near these narrow resonances is,
however, non-trivial since it requires excellent magnetic field
stability. In the above example, a magnetic field precision of 1~mG
is required to reach $A=1000$~$a_B$ and a field instability and
inhomogeneity of $0.01$~mG are necessary to determine $A$ to below
$1\%$. Since an instability of 1~mG has been achieved near a narrow
resonance \cite{mark05} without magnetic field shielding, we are
optimistic that 0.01~mG instability can be achieved. If so, a $1\%$
measurement on the scattering length can be translated into a
sensitive test of the mass ratio variation on the level of $|\delta
\beta/\beta| \sim 10^{-11}$.

Another possibility to obtain large enhancement factors is by
optical Feshbach resonances \cite{fedichev, theis05}. In this case,
the coupling strength can be externally controlled, and the bound
state can be selected for higher sensitivity on the mass ratio and
very weak dependence on the magnetic field. Using the two-photon
version of the optical Feshbach resonance for higher frequency
resolution (100~Hz frequency resolution was reported in
Ref.~\cite{wynar2000}), we expect a coupling strength of $\Delta
E=$h~5~Hz can be achieved, which, in the case of cesium, can provide
a large enhancement factor of $K\sim 10^{12}$ and a test of $\beta$
on the $10^{-14}$ level. Furthermore, the optical method allows one
to probe molecular states with different dependence on $\beta$. This
can provide significant reduction of systematic effects.

We thank Paul Julienne, David deMille and Jun Ye for stimulating
discussions. V.F. is supported by the Australian Research Council.
C.C. acknowledges support from the Chicago Material Research Center.

\end{document}